# Electron-Doped $Sr_2IrO_{4-\delta}$ ($0 \leq \delta \leq 0.04$):

## Evolution of a Disordered $J_{eff} = 1/2$ Mott Insulator into an Exotic Metallic State

O.B. Korneta[1,2], Tongfei Qi[1,2], S. Chikara[1,2], S. Parkin[1,3] L. E. De Long[1,2], P. Schlottmann[4] and G. Cao[1,2*]

[1]Center for Advanced Materials, University of Kentucky, Lexington, KY 40506

[2]Department of Physics and Astronomy, University of Kentucky, Lexington, KY 40506

[3]Department of Chemistry, University of Kentucky, Lexington, KY 40506

[4]Department of Physics, Florida State University, FL 32306

Stoichiometric $Sr_2IrO_4$ is a ferromagnetic $J_{eff} = ½$ Mott insulator driven by strong spin-orbit coupling. Introduction of very dilute oxygen vacancies into single-crystal $Sr_2IrO_{4-\delta}$ with $\delta \leq 0.04$ leads to significant changes in lattice parameters and an insulator-to-metal transition at $T_{MI} = 105$ K. The highly anisotropic electrical resistivity of the low-temperature metallic state for $\delta \approx 0.04$ exhibits anomalous properties characterized by non-Ohmic behavior and an abrupt current-induced transition in the resistivity at $T^* = 52$ K, which separates two regimes of resisitive switching in the nonlinear I-V characteristics. The novel behavior illustrates an exotic ground state and constitutes a new paradigm for devices structures in which electrical resistivity is manipulated via low-level current densities ~ 10 mA/cm$^2$ (compared to higher spin-torque currents ~ $10^7$-$10^8$ A/cm$^2$) or magnetic inductions ~ 0.1-1.0 T.

**Introduction**

It is commonly expected that iridates are more metallic and less magnetic than their 3d and 4d counterparts. The extended nature of 5d orbitals leads to a broad 5d bandwidth and a reduced Coulomb interaction U, such that the Stoner criterion anticipates a metallic, paramagnetic state. In marked contrast, many iridates are magnetic insulators (e.g., $BaIrO_3$, $Sr_2IrO_4$ and $Sr_3Ir_2O_7$), while the correlated metal $SrIrO_3$ is an exception **[1-12]**. The unusual properties of $Sr_2IrO_4$ have been attributed to the interplay of a strong spin-orbit interaction with a comparable crystalline electric field splitting **[2-5]**, which leads to a novel $J_{eff} = 1/2$ Mott state **[6,7]** associated with a topological phase **[9]**. $Sr_2IrO_4$ is a weak ferromagnet (FM) ($T_C$ = 240 K) with a saturation moment no greater than 0.14 $\mu_B$/Ir **[5,8]**. The competition between magnetic exchange interactions and lattice distortions in $Sr_2IrO_4$ gives rise to a giant magneto-dielectric shift (of order 100% in magnetic inductions B ≤ 1 T) that is driven by spin-orbit coupling rather than magnetization **[8]**.

Here, we report that introduction of very dilute oxygen vacancies into single-crystal $Sr_2IrO_{4-\delta}$ drives the following intriguing phenomena:

1) An insulator-to-metal transition signaled by a highly anisotropic resistivity that continues to decrease by ***several orders of magnitude*** below a temperature $T_{MI}$ (≥ 105 K for $\delta \leq$ 0.04), ***without saturation to a residual limit*** at the lowest temperature studied (T = 1.8 K).

2) Non-linear I-V behavior occurs with switching at three modest current thresholds (~ 10 A·cm$^{-2}$).

3) An abrupt current-induced transition in the resistivity at T* = 52 K separates two temperature regimes with different non-linear I-V characteristics.

The properties of this exotic electronic state offer a new paradigm for devices that can be manipulated by only modest electrical current densities.

**Experimental Methods**

Synthesis of flux-grown single-crystal $Sr_2IrO_4$ is described elsewhere [**5, 8,11,12**]. The crystal structures of both stoichiometric $Sr_2IrO_4$ and non-stoichiometric $Sr_2IrO_{4-\delta}$ ($\delta \leq 0.04$) single crystals were determined at T = 90 K and 295 K using a Nonius Kappa CCD X-Ray Diffractometer and Mo K$\alpha$ radiation. Chemical compositions of the single crystals were determined using energy dispersive X-ray analysis (EDX). Unlike cuprate single crystals that require a complex post-annealing sequence and a long diffusion time (e.g., days or even weeks) to significantly alter physical properties, single-crystal samples of the layered iridates undergo *significant structural alterations and radical changes in bulk transport and magnetic properties* after merely hours of post annealing. Reduced oxygen content ($0 \leq \delta \leq 0.04$) was generated by firing an as-grown single-crystal $Sr_2IrO_4$ in a TGA (Mettler-Toledo Model TGA/DSC 1) or an evacuated quartz tube at 600 $^o$C for different periods of time, depending on the composition $Sr_2IrO_{4-\delta}$ sought. (We note that structural and physical properties of other iridates such as $Sr_3Ir_2O_7$ and $BaIrO_3$ show the same sensitivity to oxygen depletion). Values of $\delta$ were determined in the TGA measurements. I-V characteristics were obtained using a Keithley 6220 current source and a Keithley 2182A nanovoltmeter. The differential resistance was measured using a built-in function of the above Keithley meters to eliminate potential thermoelectric voltages on the contact leads. Comparisons between the differential resistance and conventional resistance revealed no differences whatsoever, indicating contact effects were negligible. Measurements of magnetization M(T,H), electrical resistivity $\rho$(T,H) and thermopower S(T) were performed using either a Quantum Design PPMS or MPMS, as described elsewhere **[8].**

**Experimental Results**

$Sr_2IrO_4$ crystallizes in a reduced tetragonal structure (space-group $I4_1/acd$) due to a rotation of the $IrO_6$-octahedra about the **c-**axis by an angle ~11° **[2-4]**. This rotation corresponds to a distorted in-plane Ir1-O2-Ir1 bond angle θ that *decreases* with decreasing temperature. Our single-crystal X-ray diffraction data confirm this trend for stoichiometric crystals (δ = 0), and also reveal that θ for δ = 0.04 *increases* slightly with decreasing temperature from ***157.028°*** at 295 K to ***157.072°*** at 90 K, and the latter angle is significantly larger than that for δ = 0, (i.e., θ = ***156.280°*** at 90 K), as shown in **Fig. 1** and **Table 1**. The increment Δθ = ***0.792°*** is large for such a small oxygen depletion. Moreover, the volume of the unit cell V for δ ≈ 0.04 contracts by an astonishing ***0.14%*** compared to that for δ = 0 (see **Table 1**). These data indicate that dilute oxygen vacancies relax θ and reverse its temperature dependence with increasing δ, while significantly reducing the structural distortion at low T. *No such changes in the lattice parameters would be observable in X-ray diffraction data should the oxygen depletion be confined to the crystal surface and not uniformly distributed within the bulk*.

**Table 1. Lattice parameters at T = 90 K for δ = 0 and 0.04**

| δ (T=90K) | a(Å) | c(Å) | V(Å3) | Ir1-O2-Ir1 bond angle θ |
|---|---|---|---|---|
| **0** | 5.4836(8) | 25.8270(5) | 776.61(22) | 156.280° |
| **0.04** | 5.4812(3) | 25.8146(16) | 775.56(8) | 157.072° |

The Ir1-O2-Ir1 bond angle θ is an important focus of this study, as it controls the hopping of the 5d electrons and superexchange interactions between Ir atoms via the bridging O sites **[10]**, and is therefore expected to influence physical properties. For example, the **a**-axis resistivity $\rho_a$ (**c**-axis resistivity $\rho_c$) is reduced by a factor of **$10^{-9}$** (**$10^{-7}$**) with doping at T = 1.8 K as δ changes from 0 to ~ 0.04 (see **Figs. 2a** and **2b**). For δ ≈ 0.04, there is a sharp insulator-to-metal transition near $T_{MI}$ = 105 K, resulting in a reduction of $\rho_a$ ($\rho_c$) by a factor of **$10^{-4}$** (**$10^{-1}$**), from just below $T_{MI}$ to T = 1.8 K (**Figs. 2b**). The strong low-T anisotropy reflected in the values of $\rho_a$ and $\rho_c$ and their sensitivity to δ are consistent with a nearly 2D, strongly correlated electron system. Below 20 K, $\rho_a$ has linear-T dependence *without saturation to a residual resistivity limit*; and although there is a plateau in $\rho_c$ for 5 < T < 35 K, it is followed by a very rapid downturn near $T_a$ = 5 K (**Fig. 2b** inset), which indicates a sudden, rapid decrease in inelastic scattering.

Oxygen depletion also changes the electronic density of states $g(E_F)$ of $Sr_2IrO_{4-\delta}$, as reflected in the thermoelectric power S(T), as shown in **Fig. 2c**. A peak in the **c**-axis $S_c(T)$ for δ = 0.04 is only ~ 1/3 of the peak value observed for δ = 0. Since S(T) measures the voltage induced by a *temperature gradient* (which cannot be confined to the surface of the crystal), the drastic changes in S(T) shown in **Fig. 2c** further reinforces our conclusion that oxygen depletion is a bulk effect, as indicated by the observed changes in lattice parameters discussed above. The strong reduction of $S_c(T)$ for δ ≈ 0.04 indicates an increase of $g(E_F)$ with increasing δ, since S ∝ $1/g(E_F)$ **[13]**. The rapid increase of $g(E_F)$ with increasing δ is also evident in **Fig. 3a**, where data for the **a**-axis resistivity of five representative single crystals of $Sr_2IrO_{4-\delta}$ document a decrease of $\rho_a$(T=1.8K) by *nine orders of magnitude* as δ changes from 0 to ~ 0.04. These rapid changes in transport properties with doping are much stronger than those observed for Lifshitz transitions in metallic alloys where the Fermi level crosses small pockets with doping or applied pressure **[14]**. On the other hand, they are reminiscent of the extreme sensitivity of "correlation-gap insulators" to dilute impurities and pressure **[15,16]**.

Magnetic correlations drive a weak FM state for $\delta = 0$ below $T_C = 240$ K, but surprisingly, no corresponding anomaly has been observed in $\rho(T)$ and $S(T)$ **[4,5,8]**. We find that the case of oxygen-depleted crystals is quite different, where changes in magnetic properties with increasing $\delta$ are modest in comparison with those in the resistivity. The magnetizations $M(T)$ for $\delta = 0$ and $\delta \approx 0.04$ measured in an applied magnetic field $\mu_o H = 0.2$ T are compared with $\rho_a$ for $\delta \approx 0.04$ in **Fig. 2b**. $T_C$ is approximately 10 K higher for $\delta \approx 0.04$ than for $\delta = 0$; and for $\delta \approx 0.04$, both $M_a$ and $M_c$ exhibit a weak, yet visible anomaly at $T_{MI}$ (marked by arrows in **Fig. 2b**). Application of a magnetic field $\mu_o H = 7$ T causes a positive **c**-axis magnetoresistance of the order of 10% (**Fig. 2c**), and a downward shift of $T_{MI}$ by 6 K, which indicates the low-T metallic state is destabilized by field. A magnetic anomaly in $M_c(T)$ near $T_{MI}$ shifts rapidly upward from 100 K for $\delta = 0$ **[8]**, to 160 K for $\delta \approx 0.04$, which coincides with a slope change in $\rho_c$ (see the dashed line in **Figs**. **3b** and **3c**). In summary, FM order at high-T is stabilized, whereas the magnetization at low-T is reduced, with increasing $\delta$.

**Discussion**

*A. Metal-Insulator Transition*

Strong crystal fields split off 5d-band states with $e_g$ symmetry in stoichiometric $Sr_2IrO_4$, and $t_{2g}$ bands arise from $J = 1/2$ and $J = 3/2$ multiplets via strong spin-orbit coupling. A weak admixture of the $e_g$ orbitals downshifts the $J = 3/2$ quadruplet from the $J = 1/2$ doublet [6,7]. An independent electron picture anticipates a metallic state, since the $Ir^{4+}$ ($5d^5$) ions provide four electrons to fill the lower $J_{eff} = 3/2$ bands, plus one electron to partially fill the $J_{eff} = 1/2$ bands. However, the $J_{eff} = 1/2$ bandwidth W (W = 0.48 eV for $\delta = 0$) is so narrow, even a modest U (~ 0.5 eV) is sufficient to induce a Mott gap $\Delta \sim 0.5$ eV in the $J_{eff} = 1/2$ band [6]. W is quite sensitive to structural alterations according to a recent first-principles calculation [10] that predicts that an increased $\theta$ should cause a broadening of the $J_{eff} = 1/2$ band and a concomitant decrease of the Mott gap by 0.13 eV, if $\theta$ increases from 157° to 170°. The observed increment $\Delta\theta = 0.792°$ does not appear nearly sufficient to produce the dramatic changes we have observed in $\rho$, and we conclude that another mechanism must be responsible for $T_{MI}$.

Removal of oxygen from $Sr_2IrO_{4-\delta}$ is expected to result in electron doping of the insulating state that is observed to be stable for $\delta = 0$. According to (LDA+SO+U) band structure calculations [10] additional electrons will occupy states in four symmetric pockets located near the M-points of the basal plane of the Brillouin zone. Each pocket has an estimated filling of 2% of the Brillouin zone for $\delta = 0.04$. The situation appears analogous to doping in strongly correlated $(La_{1-x}Sr_x)_2CuO_4$ (LSCO) (where pockets of similar shape arise at the same positions in the Brillouin zone) and $La_2CuO_{4+\delta}$ [17]. There are, however, two fundamental differences: while in $Sr_2IrO_{4-\delta}$ we dope electrons, the added carriers are holes in the cuprates; moreover, $Sr_2IrO_4$ is a weak ferromagnet rather than a simple antiferromagnet, as is $La_2CuO_4$.

Oxygen depletion also introduces disorder, which is expected to lead to localization of states close to the band-edge in a quasi-2D system [18]. We assume that the Fermi level lies below the mobility edge for $\delta \neq 0$, and hence, the occupied states are all localized at high T, where the compound is a paramagnetic insulator. As T is lowered, the compound develops increasing FM polarization below $T_C$ and the intersection of the Fermi level with the majority spin band is gradually pushed closer to the mobility edge as the exchange splitting of the band increases below $T_C$. Eventually the Fermi level crosses the mobility edge, leading to metallic behavior below $T_{MI}$. The minority-spin carriers are always localized. $T_{MI}$ is clearly visible only for $\delta \approx$ 0.04 in **Fig. 3a**; evidently the electron density is not high enough to implement the crossing for smaller doping. Note that this scenario requires that $T_{MI}$ is considerably lower than $T_C$ since the electrons in the M-pockets first have to be polarized.

The reversed trend of the temperature dependence of $\theta$ and the increase of $\theta$ with doping could be consequences of increased screening in the metallic state. Furthermore, we expect the metallic state to be inhomogeneous and conductivity increases to arise from the growth and percolation of metallic patches; i.e., the metallic state develops out of a phase separation of competing states. The importance of disorder in the physical properties of $Sr_2IrO_{4-\delta}$ is corroborated by fits (**Fig. 3c** inset) using a variable range hopping (VRH) relation $\rho_a(T) = A\exp(T_o/T)^\upsilon$ with $\upsilon = ¼$ for $187 < T < 350$ K, and with $T_o$ a characteristic temperature; such behavior is also observed in nonmetallic samples **[5]**.

## B. Non-Ohmic Behavior

The exotic nature of the metallic state of $Sr_2IrO_{4-\delta}$ is revealed in striking non-Ohmic behavior, as exhibited by $\rho_c(T)$ for various applied currents I (see **Fig. 4a**; $\rho_a$ behaves similarly, and is not shown). $\rho_c$ changes slightly when $I \leq 1$ mA, but more dramatically when $I \geq 5$ mA (~ 10 A/cm$^2$ current density). Moreover, there is a characteristic temperature $T^* = 52$ K at which $\rho_c$ sharply drops for I = 5 mA, but rises for I = 14 mA (see **Figs. 4a** and **4c**), indicating a current-induced phase transition. Interestingly, the distinct downturn in $\rho_c$ below $T_a \approx 5$ K disappears for $I \geq 1$ mA, although it is insensitive to applied magnetic field (see **Fig. 3c**).

The non-linear I-V characteristic shown in **Fig. 4b** shows switching occurs at multiple threshold potentials as I varies from 0.1 µA to 50 mA. We infer a temperature T* that separates two different regions: For T < T*, there are three threshold potentials, $V_{th1}$, $V_{th2}$ and $V_{th3}$. The initial linearity in the I-V curve persists up to I = 4 mA for V < $V_{th1}$ (= 0.011 V at 5 K); between $V_{th1}$ and $V_{th2}$ linearity is briefly restored with a reduced slope. With further increases in I, the I-V response exhibits a third threshold $V_{th3}$, which marks the onset of *current-controlled negative differential resistivity* (NDR), where V across the crystal decreases as I increases. The qualitative difference between the two regions separated by T* is clearly revealed in the temperature dependences of $V_{th1}$, $V_{th2}$ and $V_{th3}$, as well as $\rho_c$, at I = 5 and 14 mA, as shown in **Fig. 4c**. Note that $V_{th2}$ and $V_{th3}$ increase with increasing T below T*. *For T > T*, the trend is reversed:* $V_{th3}$ and $V_{th2}$ shift to lower values, and $V_{th1}$ tends to zero with increasing T. Note that T* remains sharply defined in $\rho_c(T)$ at 52 K, independent of I (**Fig. 4c).** This completely rules out the possibility that self-heating plays a role in the non-Ohmic behavior.

It is noteworthy that non-Ohmic behavior or NDR, which is not commonplace for bulk materials, has been observed in the *insulating state of layered iridates*, such as stoichiometric $Sr_2IrO_4$ **[5]** and $BaIrO_3$ **[1]**, which exhibit switching behavior at *a single threshold* $V_{th}$, depending on temperature, much higher than the upper threshold $V_{th3}$ in $\delta \approx 0.04$. It was attributed to collective charge density wave (CDW) dynamics in the presence of disorder commonly seen in the CDW state **[1, 5]**. Indeed, density waves pinned by oxygen vacancies and then depinned by applied potential is one possible mechanism for the non-Ohmic behavior. The pockets at the M-points are predicted to have an elongated ellipsoidal shape in the plane, and due to the quasi-2D nature of the compound, so that a nesting condition between the pockets cannot be ruled out. Such nesting could give rise to either spin- or charge-density waves that would have to coexist in the presence of ferromagnetism.

The observed non-Ohmic behavior of the *metallic state* therefore poses intriguing questions concerning its origin: Does non-Ohmic behavior observed in both the insulating and metallic states of iridates arise from the same mechanism--- i.e., a CDW state? Note that the voltage thresholds (~ $10^{-2}$ V) observed for $\delta = 0.04$ are two orders of magnitude smaller than those (~ 1 V) observed in CDW depinning experiments **[19]**, which would suggest an extremely weak CDW pinning by defects that has not been reported before. Otherwise, the observed non-Ohmic behavior of the metallic state must signal a novel metallic state that does not follow Ohm's law.

The novel behavior of $Sr_2IrO_{4-\delta}$ clearly illustrates an intrinsically unstable ground state that readily swings between highly insulating ($10^5$ Ω cm) and metallic ($10^{-5}$ Ω cm) states via only very slight changes in oxygen content. The non-Ohmic behavior, whose origin is yet to be fully understood, constitutes a new paradigm for device structures in which resistivity can be manipulated with modest applied currents rather than large magnetic fields or much larger voltages.

GC is thankful to Drs M. Whangbo and J.W. Brill for useful discussions. This work was supported by NSF through grants DMR-0552267, DMR-0856234 and EPS-0814194 (GC), and by DoE through grants DE-FG02-97ER45653 (LED) and DE-FG02-98ER45707 (PS).

*Corresponding author; email: cao@uky.edu

**Figure Captions**

**Fig.1.** A schematic of the Ir1-O2-Ir1 bond angle $\theta$ at T = 90 K for **(a)** $\delta = 0$ and **(b)** $\delta \sim 0.04$ (the changes in $\theta$ are exaggerated for clarity); a schematic of the band structure corresponding to **(c)** $\delta = 0$ and **(d)** $\delta \sim 0.04$. Note that $\theta$ decreases for $\delta = 0$ but slightly increases for $\delta \sim 0.04$ as temperature is lowered.

**Fig.2.** The a- and c-axis resistivity, $\rho_a$ and $\rho_c$, as a function of temperature for **(a)** $\delta = 0$ and **(b)** $\delta \sim 0.04$; **(c)** c-axis thermoelectric power, $S_c(T)$, for $\delta = 0$ and $\delta \sim 0.04$. **Inset** in (b): $\rho_a$ and $\rho_c$ vs. T for 1.7 K < T $\leq$ 20 K; note a downturn near 5 K in $\rho_c$ and linear temperature dependence in $\rho_a$.

**Fig.3.** Temperature dependence of **(a)** log $\rho_a$ for several representative values of $\delta$, **(b)** the a- and c-axis magnetization, $M_a$ and $M_c$, for $\delta = 0$ (thin lines) $\delta \sim 0.04$ (thick lines) and $\rho_a$ (right scale) for $\delta \sim 0.04$, and **(c)** $\rho_c$ at 0 and 7 T. **Inset** in (c): log $\rho_a$ vs. $T^{-1/4}$.

**Fig.4.** **(a)** Temperature dependence of $\rho_c$ for several representative values of current I; **(b)** I-V curves at several representative temperatures and **(c)** temperature dependence of $\rho_c$ at I = 5 and 14 mA and threshold $V_{th}$ (right scale).

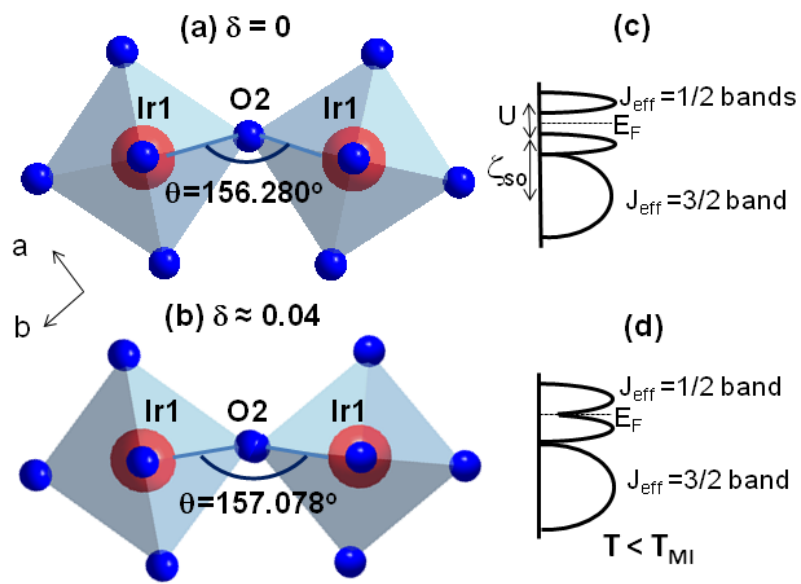

**Fig. 1**

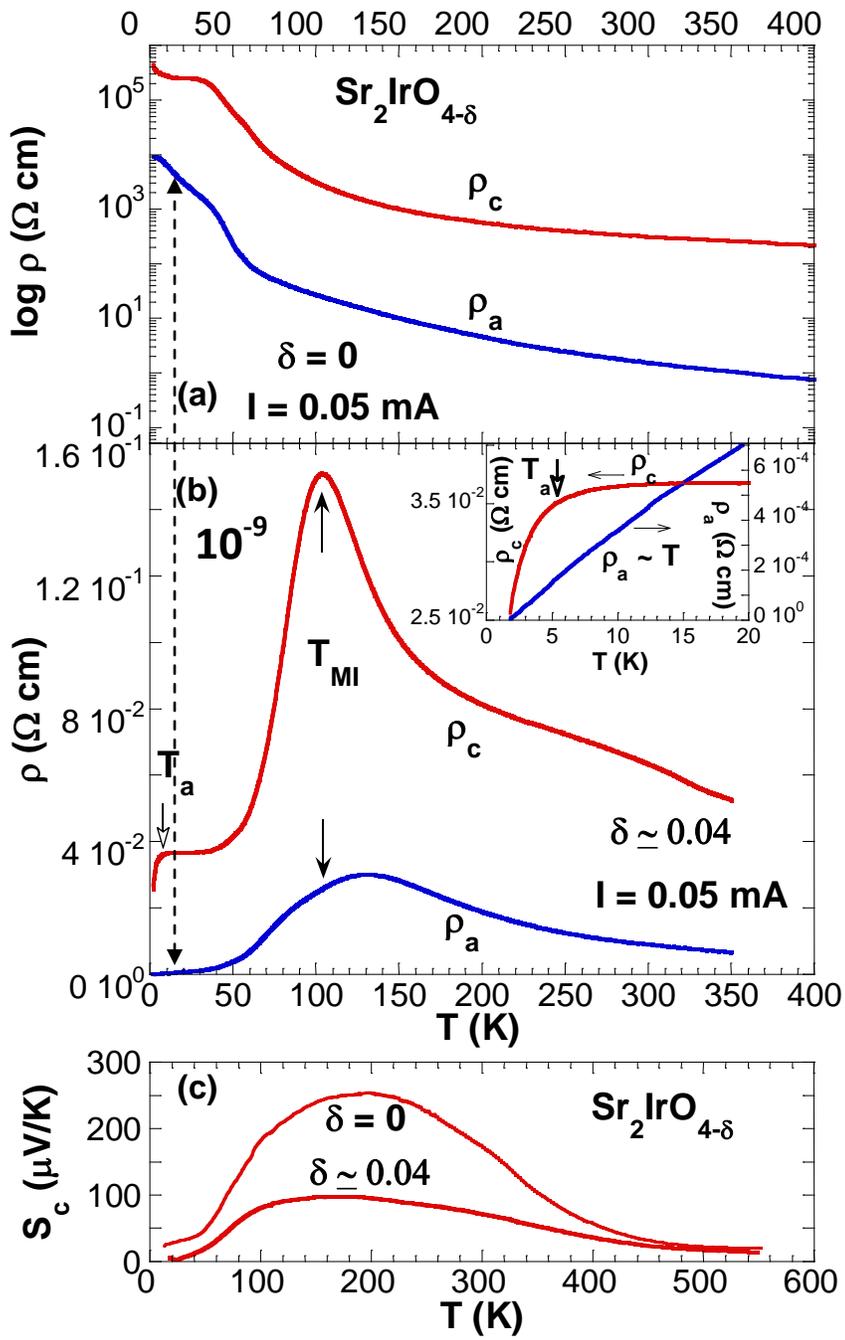

Fig. 2

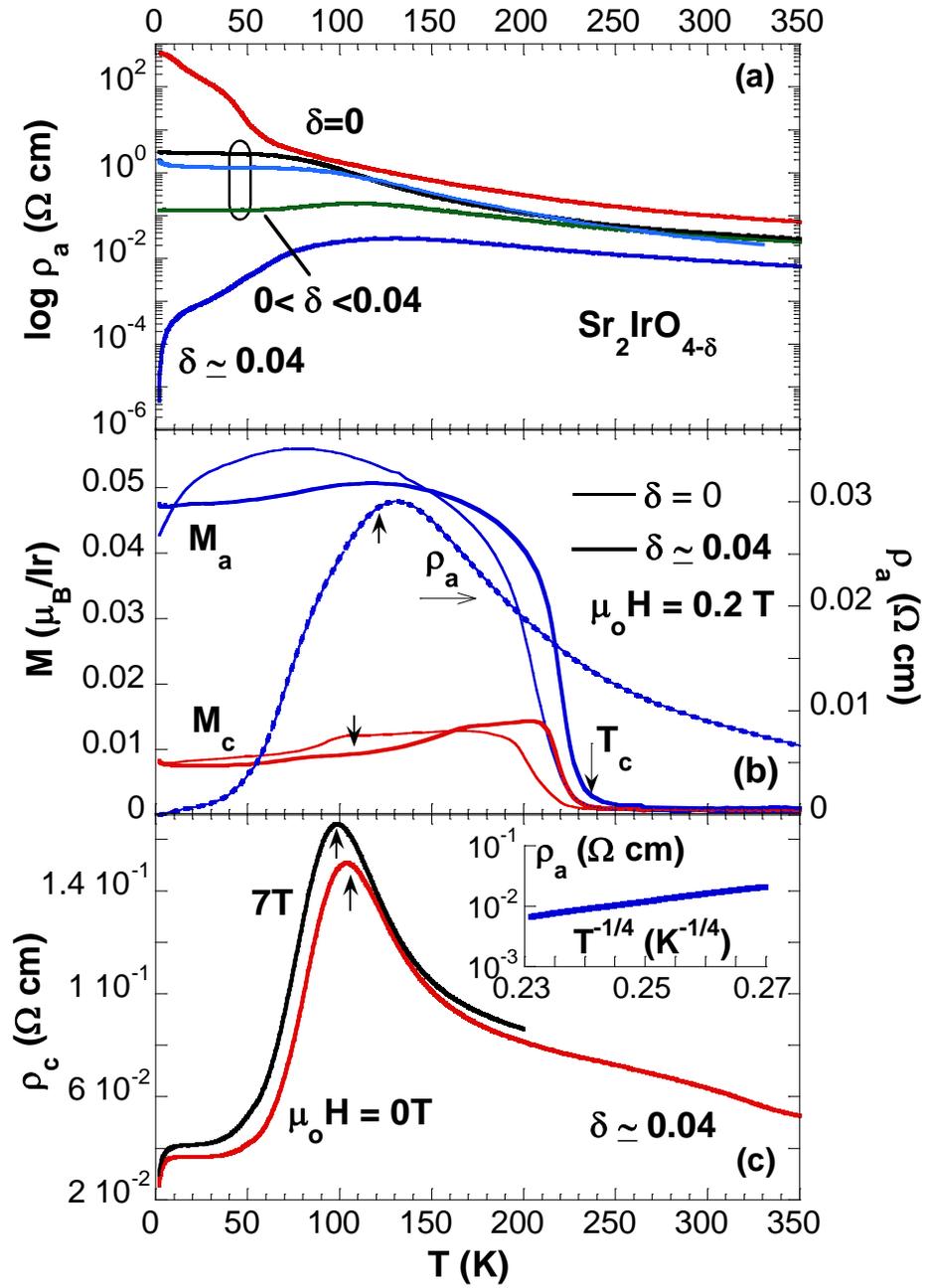

Fig.3

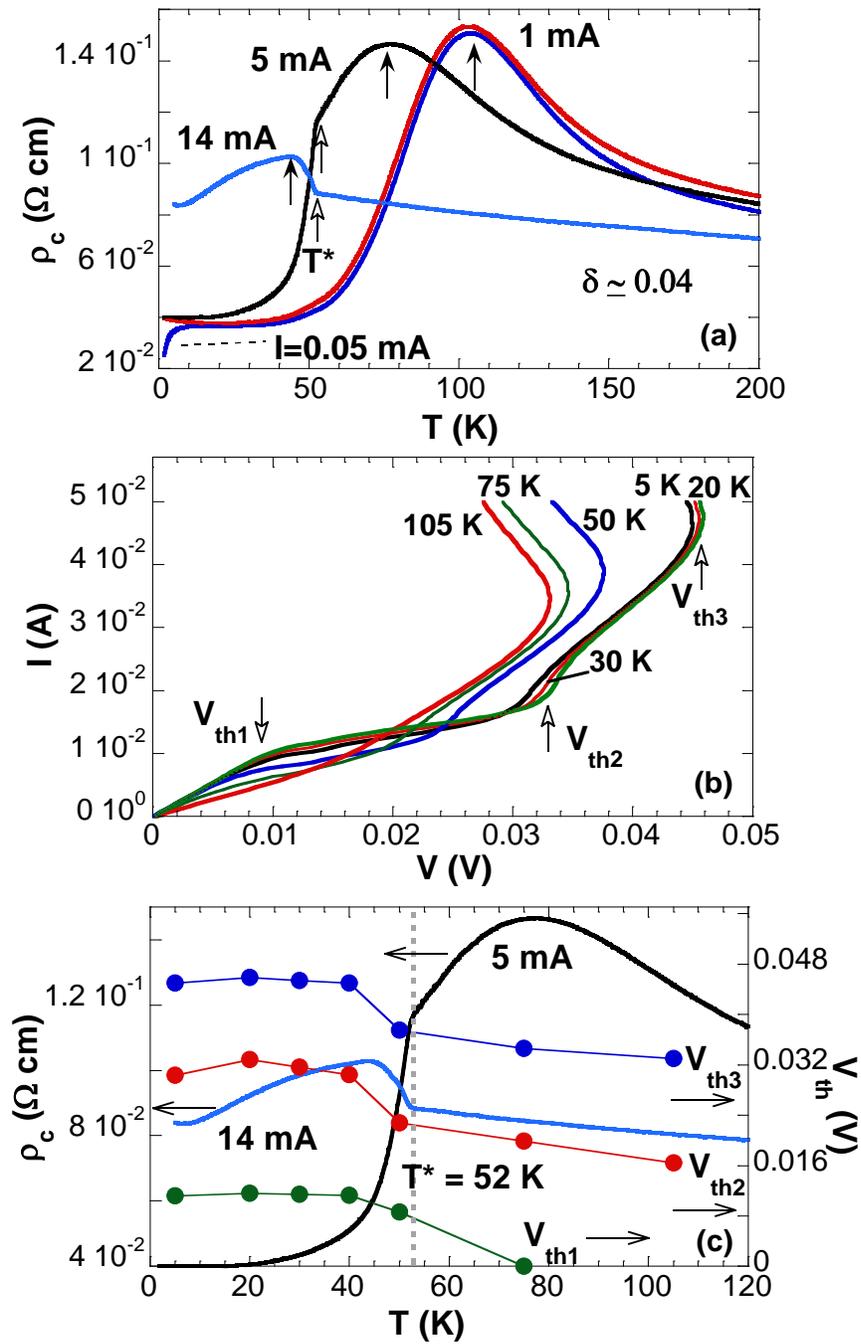

Fig.4